# Unleashing the Potential of LTE for Next Generation Railway Communications

P. Fraga-Lamas, J. Rodríguez-Piñeiro, J. A. García-Naya, and L. Castedo

Department of Electronics and Systems, University of A Coruña, Spain

**Abstract.** In an increasingly demanding marketplace that will put great strain on railway services, research on broadband wireless communication must continue to strive for improvement. Based on the mature narrowband GSM technology, Global System for Mobile Communications-Railways (GSM-R) has been deployed both for operational and voice communications. Although GSM-R fulfills the requirements of current railway services, it imposes limited capacity and high costs that restrict enhancements of operational efficiency, passenger security and transport quality. 4G Long Term Evolution (LTE) is expected to be the natural successor of GSM-R not only for its technical advantages and increasing performance, but also due to the current evolution of general-purpose communication systems. This paper examines the key features of LTE as well as its technical ability to support both the migration of current railway services and the provisioning of future ones.

**Keywords:** Railway applications, quality of service, LTE for railway, all-IP, VoLTE, interoperability, ProSe, GCSE, GSM-R.

## 1 Introduction

High-speed railway networks are extremely complex scenarios that have promoted many research initiatives primarily aiming to foster transportation quality. One of the strategic goals in high-speed rails focuses on the introduction of advanced broadband communication technologies allowing for improved services and coping with market needs in a rapidly changing landscape. Current railway communication technology was built in the beginning of the nineties considering readily available and well established mobile communication standards with great potential to fulfill the requirements of railway services at that time [1]. After a preliminary study on the usability of either Trans European Trunked RAdio (TETRA) or Global System for Mobile Communications (GSM), the latter was chosen because it was a proven technology in commercial use. Indeed, GSM Release 99 was standardized by European Telecommunications Standards Institute (ETSI) and it was well supported by its supplier association, the GSM Association (GSMA) Group. After extensive studies, GSM-R was finally standardized by the Union Internationale des Chemins de Fer (UIC) and the European Railways. The European Integrated Railway Radio Enhanced NEtwork (EIRENE) project was launched in 1992 as an alliance between ETSI, railway operators and telecommunications manufacturers. EIRENE's aim was to specify the functional and technical requirements for railway mobile networks. Two leading working groups were established within EIRENE for this task: a functional group and a project team. The functional group defined the Functional Requirements Specification (FRS) which mainly describes the mandatory features to ensure interoperability across borders. The project team developed the System Requirements Specification (SRS) based on the functional requirements. This document defines the technical characteristics related to railway operation, thus identifying and specifying the additional Advanced Speech Call Items (ASCI) features [2].

A first draft of these EIRENE specifications was finalized in 1995 when the Mobile Radio for Railway Networks in Europe (MORANE) project was launched with the involvement of the UIC; the major railways in France, Italy and Germany; the European Commission, and a limited number of GSM suppliers. The aim was to specify, develop, test and validate prototypes of a new radio system, meeting both functional and system requirements specifications. In 1997, the UIC prepared a Memorandum of Understanding (MoU) to compel railway companies to only invest and cooperate in the implementation of GSM-R. This MoU was signed in 1998 by 32 railways all over Europe increasing up to 37 in 2009, including railways outside Europe. An Agreement on Implementation (AoI) came into effect in 2000 where the 17 signing railway companies have stated their intention to begin national GSM-R implementation by 2003 at the latest. From then on, GSM-R became the railway technology until now, when the rapid pace of commercial technologies are the driving force for further research on alternatives like LTE.

This paper provides an understanding of the progress of mobile technologies in railway domain since GSM-R. It describes the motivations for the different alternatives over time and the evolution of the railway requirements with its main specifications and recommendations. The aim of the study is to envision the potential contribution of LTE to provide additional features that GSM-R could never support.

The remainder of this article is structured as follows. Section 2 reviews GSM-R services. The aim is to identify what is required to roll-out LTE to address specific requirements for railway communication services. Next, Section 3 shows an overview of GSM-R shortcomings and LTE advantages. Current status of standardization is detailed in Section 4 in order to understand the evolution of the involved requirements and technologies. In Section 5, taking both the services and the associated operational requirements as a starting point, a formal analysis is introduced to study the feasibility of LTE for next generation railway networks. Next, the strategic roadmap to ensure a smooth migration is described in Section 6. Finally, the last section is devoted to the conclusions and future research lines.

## 2   GSM-R: Railway-Specific Services and Requirements

Based on GSM Phase 2 and Phase 2+ recommendations, GSM-R was analyzed to provide maximum redundancy and achieve maximum system availability. GSM-R provides two fundamental services: voice communication and transmission of European Train Control System (ETCS) messages. Both GSM-R and ETCS constitute the European Rail Traffic Management System (ERTMS) [3]. The UIC initiated the so-called ERTMS/GSM-R project to bring together existing and future implementers. Furthermore, ERTMS/GSM-R manages the UIC roll-out plan aiming to update the existing specifications of GSM-R. This common development has continued up to now, maintaining close cooperation with ETSI and the GSM-R industry.

The FRS version 7.4.0 [4] and SRS version 15.4.0 [5], designated as European Railway Agency (ERA) GSM-R Baseline 0 Release 4, were published in 2014 and represent the latest specifications. Such documents involve the description of mandatory requirements relevant to interoperability of the rail system within the European Community, according to the Directive 2008/57/EC [6], incorporating requirements for a major milestone towards an IP-based core network architecture [7].

The areas covered in the EIRENE SRS can be outlined as follows:

- GSM-R network configuration, applicable to ER-GSM band frequencies, provides a guidance to meet performance levels, GSM-R coverage, speed limitations, handover and cell selection, and call set-up time requirements. Broadcast and group call areas are also defined. The level of coverage should be at least 95% of the time over 95% of the designated coverage area for a radio installed in a vehicle with an external antenna.
- Mobile equipment specification distinguishes five types of mobile radios: Cab radio and the Human-Machine Interface (HMI) for transmission of voice and non-safety data; EIRENE-compliant general purpose radio; EIRENE-compliant operational radio with functions to support railway operations; shunting radio; and ETCS data-only radios.
- EIRENE numbering plan requirements and constraints, call routing and structure of Functional Numbers.
- Subscription management which handles the requirements for call priorities, encryption and authentication, broadcasts and Closed User Groups (CUGs).
- GSM-R operation modes: high-priority voice calls for operational emergencies (railway emergency calls); shunting mode, including the definition of user privileges; and an optional direct-mode communication providing short range fall-back communications between drivers and track-side personnel.

Some requirements are defined by individual railway companies [8]:

- Fixed network elements (links, switches, terminal equipment, etc) and their specifications with respect to Reliability, Availability, Maintainability and Safety (RAMS) (EN50126, EN50128, EN50129), network interconnections and capacity. The fixed network must also support a specified set of services to provide end-to-end functionality. The inter-working between the fixed and mobile side of the network must also be considered.
– Requirements for signaling systems to be used within the fixed network.
– Non-mandatory specifications of controller equipment are provided by FRS, although details of such equipment and the interface between the equipment and the GSM-R network are assigned to the railway operator.
– System management functionality and platforms; in particular, the specification of fault, configuration, accounting, performance and security management requires various type of approvals to allow equipment to be connected to the network, i.e. safety approvals for each railway.
– Roaming on to a national public GSM network as part of a disaster recovery strategy in case of a loss of service.

**Table 1. Voice telephony services to be supported.**

| Voice-Call / Radio type | Cab | ETCS data only | General purpose | Operational | Shunting |
|---|---|---|---|---|---|
| Point-to-point | MI | NA | M | M | M |
| Public emergency | M | NA | M | M | M |
| Broadcast | M | NA | M | M | M |
| Group | MI | NA | M | M | M |
| Multi-party | MI | NA | O | O | M |

**Table 2. Data services to be supported.**

| Data / Radio type | Cab | ETCS data only | General purpose | Operational | Shunting |
|---|---|---|---|---|---|
| Text message | MI | NA | M | M | M |
| General data applications | M | O | O | O | O |
| Automatic fax | O | NA | O | O | O |
| ETCS train control | NA | MI | NA | NA | NA |

According to the last EIRENE specifications, the railway integrated wireless network should meet the following general and functional requirements under the categories: Mandatory for Interoperability (MI), Mandatory for the System (M), Optional (O) or Not Applicable (NA), depending of the type of radio.

– Services: voice (Table 1), data (Table 2) and call related features. The required call set-up times shown in Table 3 shall be achieved for interoperability (MI) in 95% of cases and for 99% of cases shall not be more than 1.5 times the required call set-up time.
– Railway EIRENE-specific applications are summarized in Table 4.
– Direct mode facility for local set-to-set operation without network infrastructure.
– Railway specific features: set-up of urgent or frequent calls through single keystroke or similar, display of functional identity of calling/called party, fast and guaranteed call set-up, seamless communication support for train speeds up to 500 km/h, automatic and manual test modes with fault indications and control over mobile network selection and system configuration.
– Dedicated buttons that allow quick access to emergency call, Push-to-talk (PTT) and support for Link Assurance Signal (LAS) are required. Latest generation railway features such as Originator-To-Dispatcher-Information (OTDI), late entry, and frequency hopping in group call shall be considered. In addition, layouts for further features are presented: enhanced Presentation of Functional Number (ePFN), Driver's Safety Device alarm, Plain Text Messages, Presentation of the Functional Number (FN) of the initiator of a Railway Emergency Call (REC) and Alerting of a Controller.

**Table 3. GSM-R Call set-up time requirements.**

| Call type | Call set-up time |
|---|---|
| Railway emergency call | <2 s (MI) |
| Group calls between drivers in the same area | <5 s (MI) |
| All operational mobile-to-fixed calls not covered by the above | <5 s (O) |
| All operational fixed-to-mobile calls not covered by the above | <7 s (O) |
| All operational mobile-to-mobile calls not covered by the above | <10 s (O) |
| All low priority calls | <10 s (O) |

**Table 4. Specific features to be supported.**

| Feature / Radio type | Cab | ETCS data only | General purpose | Operational | Shunting |
|---|---|---|---|---|---|
| Functional addressing | MI | NA | M | M | M |
| Location dependent addressing | MI | M | O | O | O |
| Direct mode | O | NA | NA | O | O |
| Shunting mode | MI | NA | NA | NA | M |
| Multiple communications within the train | MI | NA | NA | NA | NA |
| Railway emergency calls | MI | NA | O | M | M |

A common minimum standard of performance is required to EIRENE compliant mobiles, although coverage and speed-limitations values are described in SRS, it should be noted that high-speed railway systems [9] (in operation, under construction and planned) have to cope with speeds of at least 250 km/h while enabling speeds over 300 km/h under appropriate circumstances. Generally, speeds around 200-220 km/h represent the threshold for upgraded conventional lines. Nevertheless stable wireless connections have to be ensured at the moving speed of 500 km/h or even more in the future [10].

Quality of Service (QoS) mechanisms shall ensure the priorization and pre-emption of critical services. Even though current wireless networks support various QoS policies depending on different traffic types, QoS for railway-critical communications and real-time applications shall be examined. QoS control is mandatory for resource management, safety, punctuality, efficiency and accident prevention of trains, to ensure immediate reaction to emergencies and on-time operations. Strict latency requirements are needed for the seamless transmission and reception of data regarding the train position and status, and Movement Authority (MA) permission between the in-service train and the control center, i.e. the transmission error probability over one train line should be less than 1% per hour and 99% of ETCS data should have a maximum latency of < 0.5 s [11, 12]. QoS parameters with its percentage of availability are shown in Table 5.